\documentclass[aps,prb,superscriptaddress,twocolumn,showpacs]{revtex4-2}
\usepackage{amssymb}
\usepackage{epsfig}
\DeclareMathAlphabet{\pazocal}{OMS}{zplm}{m}{n}            
\usepackage{graphicx}
\usepackage[usenames, dvipsnames]{xcolor}
\usepackage{bm}
\usepackage[normalem]{ulem}     
\usepackage{soul}
\usepackage{amsmath}
\usepackage{braket} 
\usepackage{rotating}
\usepackage{multirow} 
\usepackage{mhchem}
\DeclareMathAlphabet{\pazocal}{OMS}{zplm}{m}{n}            
\usepackage{graphicx}
\usepackage{hyperref}
\usepackage[normalem]{ulem}
\usepackage{braket} 
\usepackage{booktabs}
\usepackage{tabularx}
\usepackage{multirow}
\usepackage{graphicx}

\newcommand\redsout{\bgroup\markoverwith{\textcolor{red}{\rule[0.5ex]{2pt}{0.4pt}}}\ULon}
\newcommand\bluesout{\bgroup\markoverwith{\textcolor{blue}{\rule[0.5ex]{2pt}{0.4pt}}}\ULon}

\newcommand{\SPhide}[1]{{}}

\begin{document}
\title{A Universal Framework for Controlling Non-Relativistic Spin Splitting\\
}        


\author{Aniruddha Ray}
\affiliation{Department of Physics, Indian Institute of Technology Bombay, Mumbai 400076, India}

\author{Subhadeep Bandyopadhyay}  
\affiliation{Consiglio Nazionale delle Ricerche (CNR-SPIN),  Unità di Ricerca presso Terzo di Chieti, c/o Università G. D'Annunzio, 66100 Chieti, Italy}

\author{Sayantika Bhowal}
\email{sbhowal@iitb.ac.in}
\affiliation{Department of Physics, Indian Institute of Technology Bombay, Mumbai 400076, India}

\date{\today}

\begin{abstract}
Non-relativistic spin splitting in antiferromagnets has recently attracted considerable attention. Here, we present a universal framework for controlling such spin splitting by identifying and manipulating the key atomic distortions that govern it through external perturbations. We demonstrate this concept by tuning the spin-splitting energy in three representative materials with diverse symmetries: inversion-symmetric MnF$_2$, ferroelectric BaCuF$_4$, and LaMnO$_3$/$R$MnO$_3$ superlattices. Our results emphasize the essential role of higher-order magnetic multipoles and the intrinsic structure–spin correlations in these systems, thereby advancing current efforts to control spin splitting in real materials and motivating future experimental studies.
\end{abstract}

\maketitle

\section{Introduction}

The discovery of lifted degeneracy of non-relativistic origin in spin-polarized bands of compensated antiferromagnets has recently attracted considerable attention, driven both by fundamental interest and potential applications in spintronics and related fields \cite{Hayami2019, Yuan2020, Yuan2021, Smejkal2022PRX, YuanZunger2023, Guo2023, Zeng2024, Lee2024PRL, Krempask2024, Reimers2024, Aoyama2024, Lin2024, Kyo-Hoon2019, Libor2020, Libor2022Review, Paul2024, Bai2024,Hayami2020PRB,Jungwirth2024,Cheong2024,Radaelli2025}. Early studies were primarily focused on uncovering the underlying symmetries responsible for such non-relativistic spin splitting (NRSS) \cite{Yuan2021, Smejkal2022PRX, Guo2023, Zeng2024,Hayami2020PRB,Cheong2025}, identifying candidate materials from the existing database, and exploring the unconventional and diverse physical phenomena that emerge as a result \cite{Naka2019, Hernandez2021, Shao2021, Bose2022, Hu2024,Bai2022, Karube2021,Bai2022, Karube2021,Mazin2022, Zhu2023, Banerjee2024, Chakraborty2024, Zhang2024, Lee2024,Zhou2024, Yershov2024,Libor2023, McClarty2024, Liu2024, Morano2024, BhowalSpaldin2024,Bhowal2025}.

A rapidly growing recent direction in the field focuses on controlling NRSS in real material systems, particularly through the application of electric fields. For example, in ferroelectric materials,  the reversal of NRSS with the reversal of polarization under an applied electric field has been demonstrated theoretically \cite{Stroppa_2025,Libor_2025_BaCuF4}. An analogous control of NRSS across an antiferroelectric –ferroelectric transition, driven by electric field, has also been predicted \cite{Duan_2025}. Multiferroics further provide a versatile platform, as their coupled magnetic and electric order parameters allow external electric fields to tune NRSS by manipulating magnetic domains \cite{Dong2025}.

Complementary to the electric field–driven control of NRSS in (anti-)ferroelectric and multiferroic materials, our recent studies \cite{Bandyopadhyay2025,BandyopadhyayPRB2025} have demonstrated that NRSS can also be tuned in inversion-symmetric systems using external perturbations such as optical excitation, strain, and chemical engineering. This naturally raises the question of whether there is a general framework for controlling NRSS in real materials.
In this work, we address this question by introducing a universal framework for controlling NRSS.

Our key idea is to identify the phonon distortion mode that governs NRSS. We emphasize that such phonon modes are universally present in materials exhibiting NRSS. However, they may couple with other structural distortion modes, which, in turn, influence the NRSS. The central strategy is, therefore, to tune the amplitude of the relevant phonon mode using suitable external perturbations. While in different material systems, different phonon modes may play the dominant role, and accordingly, the external perturbation required to control them can also vary from one material to another, the framework to control NRSS by modulating a particular phonon mode using external perturbation remains general. 

We demonstrate our key idea using a combined structural and multipolar analysis. The analysis of structural phonon modes is essential to understand the distortions present in a crystal structure, while the multipolar analysis provides a framework for pinpointing the phonon modes most relevant to NRSS. Previous studies have shown that the ferroically ordered magnetic octupoles are correlated with the NRSS of $d$-wave symmetry \cite{BhowalSpaldin2024}. Importantly, not only the presence or absence of these magnetic octupoles, but also their magnitude correlates directly with the strength of NRSS \cite{BhowalSpaldin2024}. Thus, identifying which structural distortion modes control the magnitude of the magnetic octupole offers a pathway to engineer and tune NRSS.

Building on these tools, we illustrate our idea by examining three representative material systems of different structural and magnetic symmetries that exhibit $d$-wave NRSS, viz., MnF$_2$, BaCuF$_4$, and LaMnO$_3$/RMnO$_3$ superlattices ($R$ = Gd, Eu, Sm, Nd). MnF$_2$ crystallizes in an inversion-symmetric rutile structure, while BaCuF$_4$ is FE. We demonstrate that, despite their contrasting symmetries, NRSS can be effectively tuned in both cases. In MnF$_2$, biaxial strain changes the $A_{1g}$ phonon distortion, leading to an increase in ferroic magnetic octupoles and, hence, the NRSS. In BaCuF$_4$, we find that CuF$_6$ octahedral rotations are coupled to polar distortions. Therefore, reversing the latter with an external electric field flips the octahedral rotation, and hence, the magnetic octupoles at Cu sites, resulting in a switching of the NRSS. On the other hand, LaMnO$_3$/RMnO$_3$ superlattices offer an intriguing case, in which, even though bulk LaMnO$_3$ is inversion symmetric, the superlattice induces uncompensated antipolar distortions from La–$R$ size mismatch, generating an electric polarization. In addition to facilitating electric-field switching of NRSS similar to BaCuF$_4$, we show that the superlattice geometry allows chemical tuning by varying the $R$-site cation, which not only changes the polarization but also modifies the coupled octahedral rotations. Consequently, magnetic octupoles also change, providing a route for engineering NRSS.

Our work contributes to the ongoing efforts \cite{Stroppa_2025,Libor_2025_BaCuF4,Duan_2025,Dong2025, Bandyopadhyay2025,BandyopadhyayPRB2025} of controlling NRSS and related effects by providing a general framework for achieving this. It also highlights the intriguing interplay between structure and magnetism present in this class of materials. The rest of the manuscript is organized as follows. In Section II, we provide the details of the computational methods used in our work, with particular emphasis on the multipolar analysis. This is followed by the results and discussion of our findings on the three materials in Section III. Finally, Section IV summarizes our main findings and highlights the broader outlook of this work.

\section{Computational Details}\label{method}
To demonstrate the three different ways of tuning NRSS in the considered example materials, we have carried out electronic structure calculations within the framework of density function theory (DFT) using the plane wave-based projector augmented wave (PAW) method \cite{Bloch1994, Kresse1999}, as implemented in the Vienna ab initio simulation package (VASP) \cite{Kresse1993, Kresse1996}. For strain engineering in MnF$_2$, we employed the exchange-correlation functional within the local density approximation (LDA) \cite{kohn1965self}, following earlier work \cite{BhowalSpaldin2024}. To account for the correlated Mn-3$d$ electrons, we included an onsite Hubbard interaction \cite{dudarev1998electron} of $U_{\rm eff}=U-J=4$ eV \cite{BhowalSpaldin2024}. An energy cutoff of 520 eV for the plane waves and a $10\times10\times14$ $k$-mesh for Brillouin zone (BZ) sampling was used to achieve electronic convergence. To apply in-plane biaxial strain to the crystal structure of MnF$_2$, the in-plane lattice parameters, $a$ and $b$, were changed to the strained values. Accordingly, the out-of-plane lattice constant $c$ was adjusted to keep the unit-cell volume unchanged. We further optimized the ionic positions of the strained structures until the Hellmann–Feynman forces on all atoms were reduced below 0.005 eV/\AA.

The electric field tuning of NRSS is demonstrated with the example material $\mathrm{BaCuF_4}$. All calculations were performed with the relaxed structure of $\mathrm{BaCuF_4}$.We carried out structural optimization of the reported $Cmc2_1$ structure \cite{DANCE_ACuF4, Claude_2006_BaMF4} by relaxing both the ionic positions and lattice parameters until the Hellmann–Feynman forces on all atoms were reduced below 0.005 eV/\AA.
The electronic structure of $\mathrm{BaCuF_4}$ was carried out within the DFT+$U$ approach, using the generalized gradient approximation (GGA) to the exchange-correlation potential and considering an onsite Coulomb interaction \cite{Liechtenstein1995} for the Cu-3$d$ electrons with $U=7$ eV and $J=$ 0.9 eV \cite{Garcia_BaCuF4_2018}. Electronic convergence was achieved with an energy cutoff of 600 eV for the plane waves and a $k$-mesh of $6\times4\times6$.

The chemical engineering of NRSS is demonstrated for the LaMnO$_3$/$R$MnO$3$ superlattice structures, where $R =$ Gd, Eu, Sm, and Nd. Both the lattice parameters and the ionic coordinates were optimized for all the superlattice structures until the Hellmann–Feynman forces on all atoms were reduced below 0.005 eV/\AA. For electronic convergence, we used an energy cutoff of 520 eV and a $k$-grid of $12\times12\times8$. The exchange–correlation functional is approximated as GGA, and a Hubbard $U$ \cite{Liechtenstein1995} is applied to the Mn-3$d$ orbitals with $U$=5 eV and $J$=1 eV \cite{BandyopadhyayPRB2025}.

For all the structures, symmetry adapted mode analyses were performed using the ISODISTORT software\cite{Isodistort, Isodistort1}. The unstrained $P4_2/mnm$ structure of MnF$_2$, the non-polar $Cmcm$ structure of BaCuF$_4$, and the cubic $Pm\overline{3}m$ structure of LaMnO$_3$ were considered as reference structures for the symmetry adapted mode analysis of the strained MnF$_2$, ferroelectric BaCuF$_4$, and the LaMnO$_3$/$R$MnO$3$ superlattices, respectively.

\subsection{Multipole Analysis}

In our current framework for controlling NRSS, the multipolar analysis serves as a guideline to identify possible engineering strategies. We note that for all the candidate $d$-wave NRSS materials, relevant to our discussion, a ferroic ordering of magnetic octupoles has previously been reported as crucial \cite{BhowalSpaldin2024,Nag2024}. In compensated AFMs with broken time-reversal ($\cal T$) symmetry, there are always some higher-order magnetic multipoles with ferroic ordering. 

Since the magnetic octupole ${\cal O}_{ijk}$ is the most relevant multipole for our present discussion, we focus only on ${\cal O}_{ijk}$. It is an inversion ($\cal I$)-symmetric but $\cal T$-broken, rank-3 magnetic multipole, defined as ${\cal O}_{ijk}=\int r_i r_j \mu_k(\vec{r}) \, d^3r$.
Physically, it represents the anisotropic part of the magnetization distribution with the specific component of ${\cal O}_{ijk}$ denoting the anisotropy on the $r_i r_j$ plane for the $k$ component of the magnetization density $\vec \mu (\vec{r})$. 
The octupole tensor can further be decomposed into totally symmetric and non-symmetric components, which we discuss later \cite{Urru2022}. 

We compute the atomic-site magnetic octupole and other relevant multipoles within an atomic sphere around each atom by decomposing the density matrix $\rho_{l_1 m_1 l_2 m_2}$, obtained from the self-consistent calculations within DFT (as discussed earlier), into irreducible (IR) spherical tensor components $ w_t^{kpr}$ \cite{Cricchio2010, Spaldin2013}. Here, $k$, $p$, and $r$ represent the spatial index, spin index, and the rank of the tensor, respectively, while $t$ labels the tensor component. The spatial index $k$ is determined by the spatial part of a given multipole, e.g., $k=2$ corresponds to the quadratic $r$ dependence of ${\cal O}_{ijk}$ tensor. The spin index $p$ can, in general, take only two values: 0 and 1, representing ${\cal T}$-even (e.g., charge) and ${\cal T}$-odd (e.g., magnetic multipoles) multipoles, respectively. Therefore, $p=1$ for magnetic octupole tensor. For given $k$ and $p$ indices, the rank $r$ can vary from $|k-p|$ to $|k+p|$. Consequently, for the magnetic octupole tensor, $r$ can take three possible values, i.e., $3, 2, 1$, with the corresponding IRs $w^{213}, w^{212}, \text{and } w^{211}$ representing the totally symmetric, the non-symmetric toroidal quadrupole moment ${\cal Q}^{(\tau)}_{ij}$), and the toroidal moment $t_i^{(\tau)}$, respectively \cite{Urru2022}. For each of these IRs of a given rank $r$, there are $(2r+1)$ components, labeled as $t=-r,\ldots,r$. Thus, the octupole components $w^{213}, w^{212}, \text{and } w^{211}$ have 7, 5, and 3 components, respectively. 

The formalism discussed here for the magnetic octupole is general and also applies to other charge and magnetic multipoles of different symmetries. For example, charge dipoles and quadrupoles are denoted by $w^{101}$ and $w^{202}$, with 3 and 5 components, respectively. For inversion-symmetry-broken multipoles (e.g., charge dipole), only those density matrix elements contribute for which $l_1+l_2$ corresponds to an odd number, whereas for inversion-symmetric multipoles (e.g., charge quadrupole, magnetic octupole, etc.), terms with $l_1+l_2$ =even contribute.

\section{Results and Discussions}

We now proceed to discuss three different ways of engineering NRSS using the multipolar framework discussed above and analyze the effects on band structure.

\subsection{Strain engineering of NRSS in centrosymmetric MnF$_2$}
\label{MnF2}

\begin{figure} 
  \centering
  
  \includegraphics[width=\columnwidth]{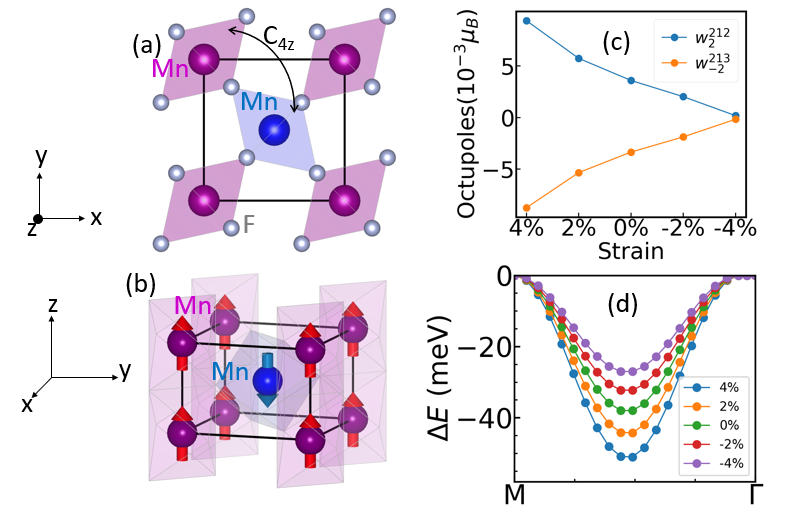}
  \caption{ Tuning of NRSS using biaxial strain. (a) Top view of the crystal structure of MnF$_2$, showing the $C_{4z}$ rotation of MnF$_6$ octahedra between the corner and the center Mn atoms. (b) Three-dimensional view of the MnF$_2$ crystal structure along with the AFM dipolar order. The red and blue arrows indicate the up and down spin polarization of the corner and central Mn ions. (c) Variation of ferroically ordered magnetic octupole components as a function of biaxial strain.  The zero value of strain indicates the unstrained structure, while the positive and negative values of strain indicate the compressive and tensile strain, respectively. (d) Energy difference between spin up and spin down band, denoted as $\Delta$E, for the top pair of valence bands along high-symmetry path, M $\rightarrow$ $\Gamma$, for various strained structures of MnF$_2$. }
  \label{fig1}
\end{figure}

The rutile structure of MnF$_2$, with space group symmetry $P4_2/mnm$, as depicted in Figs. \ref{fig1}(a) and (b), is inversion symmetric. As shown in Fig. \ref{fig1}(a), the MnF$_6$ octahedral environments around the corner and central Mn ions are connected by $C_{4z}$ rotational symmetry, which plays a crucial role in dictating the NRSS in the band structure of MnF$_2$. The ground state of MnF$_2$ has an antiferromagnetic (AFM) spin configuration, with the Mn spin moments oriented along the [001] direction, as shown in Fig. \ref{fig1}(b) \cite{deHaas1940,Seehra1984,Yamani2010}. Because of the rotated MnF$_6$ octahedra, the AFM configuration of MnF$_2$ breaks the {\it global} ${\cal T}$ symmetry, i.e., time-reversal plus translation is not a symmetry of the magnetic unit cell of MnF$_2$. This results in a spin splitting along $\Gamma \rightarrow M$ direction in the BZ of MnF$_2$ even in the absence of SOC, leading to a $d$-wave symmetry \cite{Yuan2020, BhowalSpaldin2024}.

The $d$-wave NRSS in MnF$_2$ can be described within the multipole framework by the presence of ferroically ordered magnetic octupole components $w^{212}_{2}$ and $w^{213}_{-2}$ \cite{BhowalSpaldin2024}.
The ferroic magnetic octupoles, in turn, are correlated with the local charge quadrupole component $w^{202}_{-2}$, which has opposite signs for the Mn ions, i.e., they are antiferroically ordered with the same pattern as the AFM dipolar order in MnF$_2$. Physically, the charge quadrupoles describe the anisotropic charge density around the Mn ions caused by the rotated MnF$_6$ octahedral networks. Building on the strong dependence of magnetic octupoles on the local MnF$_6$ environment, we recently proposed a strategy to tune NRSS in MnF$_2$ by jointly controlling two coupled phonon modes, leading to a transition from an NRSS AFM state to a conventional AFM state without spin splitting \cite{Bandyopadhyay2025}. In practice, such joint control of phonon modes can be achieved using state-of-the-art optical excitation techniques. In this work, we demonstrate an even simpler and more practical alternative, viz., tuning NRSS in MnF$_2$ through the application of biaxial strain.

\begin{figure}
    \centering
    \includegraphics[width=1\linewidth]{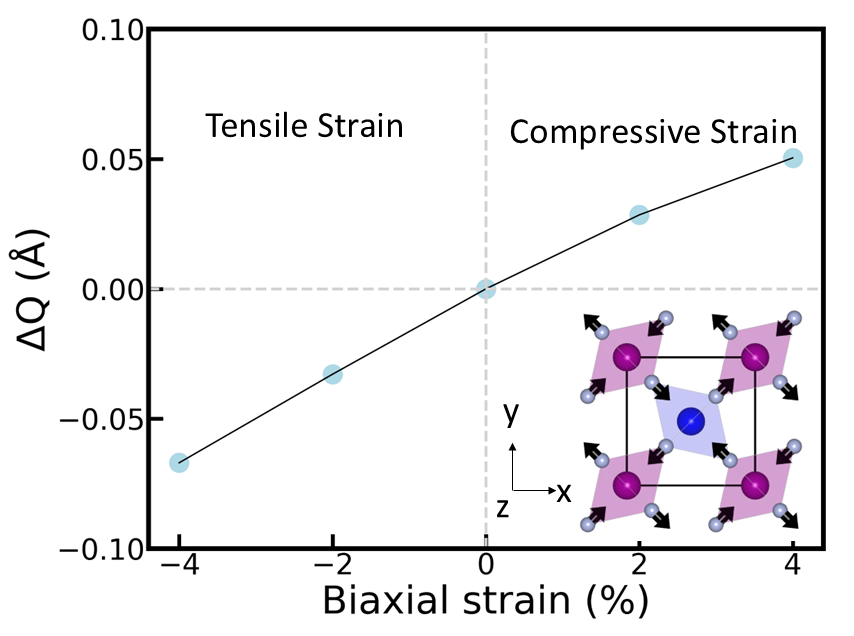}
    \caption{Variation in the amplitude of the A$_{1g}$ phonon distortion with respect to its value in the ground-state structure of MnF$_2$, denoted by $\Delta Q$, as a function of biaxial strain. The dashed line refers to $\Delta Q=0$ corresponding to the unstrained structure. Negative $\Delta Q $ refers to distortion in the reverse direction. The inset shows the $A_{1g}$ atomic distortion with the black arrows indicating the displacements of F ions. } 
    \label{fig2}
\end{figure}

From a detailed analysis of the phonon band structure of MnF$_2$, as described in Appendix \ref{appendix1}, we identify the $A_{1g}$ phonon distortion mode to be crucial for controlling NRSS. It is a stable $\Gamma$-point phonon mode \cite{Bandyopadhyay2025} of the ground-state $P4_2/mnm$ structure and lies at a frequency of 325 cm$^{-1}$. As shown in the inset of Fig. \ref{fig2}, it describes the displacement of the F atoms in the $a$-$b$ plane, thereby controlling the in-plane MnF$_6$ octahedral distortions. Since the magnetic octupoles at Mn sites in MnF$_2$ correlate with the MnF$_6$ octahedral distortions, we anticipate that by controlling the amplitude of the $A_{1g}$ phonon mode distortion, it is possible to control the magnitude of the magnetic octupoles and, hence, the NRSS.

To proceed further with this idea, we need to identify an external stimulus that can control the amplitude of the $A_{1g}$ distortion. One possible way to achieve this is through the application of biaxial in-plane strain. Since MnF$_2$ crystallizes in a tetragonal structure with $a=b\ne c$, the application of in-plane strain keeps the symmetry of the structure unchanged while affecting the Jahn-Teller-like $A_{1g}$ phonon distortion of the MnF$_6$ octahedra \cite{Bandyopadhyay2025,TMO_subhadeep}. This is also evident from our explicit calculations of the relative amplitude ($\Delta Q$) of the $A_{1g}$ distortion mode of the strained structure with respect to the unstrained structure, as depicted in Fig. \ref{fig2}. Here, compressive strain refers to positive values of strain, which contract the in-plane lattice parameters, while the tensile strain elongates them and is represented by a negative sign. As seen from Fig. \ref{fig2}, the amplitude of the distortion increases with increasing magnitude of compressive strain.  Note that since the $A_{1g}$ distortion of the unstrained structure is chosen as our reference, the relative amplitude of distortion $\Delta Q =0$ in the absence of strain.

After confirming the possibility of modulating the $A_{1g}$ distortion using biaxial strain, we now discuss the effect of biaxial strain on the ferroically ordered magnetic octupoles and the resulting effect on the band structure in the absence of SOC. Our calculations show that even in the presence of biaxial strain, the octupole components $w^{212}_{2}$ and $w^{213}_{-2}$ remain ferroically ordered, consistent with the unchanged symmetry of the structure in the presence of biaxial strain. The variation of the computed magnetic octupole components for different values of biaxial strain is shown in Fig. \ref{fig1}(c). As seen from the figure, the magnitudes of ferroic magnetic octupole components increase with increasing compressive strain, i.e., also with an increase in $A_{1g}$ distortion. Consequently, we find changes in the magnitude of the energy splitting between up- and down-spin-polarized bands, $\Delta E = E_{\uparrow} - E_{\downarrow}$. As shown in Fig. \ref{fig1}(d), consistent with our computed magnetic octupoles, our results show that $\Delta E$ also increases with increasing compressive biaxial strain, demonstrating the strain-driven tuning of NRSS in MnF$_2$.

\subsection{Electric field Tuning of NRSS in Ferroelectric BaCuF$_4$}

\begin{figure*}
    \centering
\includegraphics[width=1\linewidth]{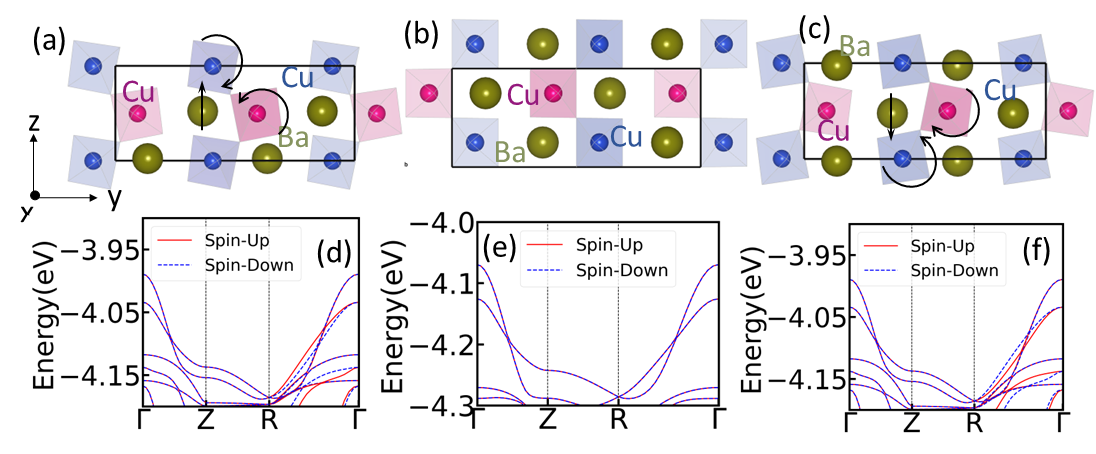} 
    \caption{Switching of NRSS in BaCuF$_4$. (a) Crystal structure ($+P$ structure) of BaCuF$_4$ with space group $Cmc2_1$. (b) Hypothetical non-polar $P=0$ structure of BaCuF$_4$ in the space group of $Cmcm$. (c) $-P$ structure of BaCuF$_4$ crystal with space group $Cmc2_1$ but opposite polar distortion. 
    Cu ions, denoted by red and blue colors in (a)-(c), have opposite spin polarization. 
    (d)-(f) Band structure of BaCuF$_4$ for the crystal structures shown in (a),(b), and (c) respectively. Absence of NRSS for the structure (b) and opposite spin splitting for structures (a) and (c) are apparent from the band structures. The zero value in energy corresponds to the Fermi energy, which is set at the minimum of the conduction band. }
    \label{fig3}
\end{figure*}

After describing the tuning of NRSS in inversion-symmetric MnF$_2$, we next consider an inversion-broken system, viz., BaCuF$_4$. BaCuF$_4$ crystallizes in the polar $Cmc2_1$ structure, which exhibits an in-phase rotation (around the $x$-axis)
of the CuF$_6$ octahedral network, accompanied by a polar displacement of the Ba atoms along $\hat z$, as depicted in Fig. \ref{fig3}(a). For an $A$-type AFM order of the Cu magnetic moments, as shown in Fig. \ref{fig3}(a), corresponding to the magnetic ordering of BaCuF$_4$ at higher temperatures \cite{DANCE_ACuF4}.
BaCuF$_4$ is known to exhibit $d$-wave NRSS in its band structure \cite{smejkal2024}. Here, we demonstrate the tuning of NRSS in BaCuF$_4$ using a combined phonon and multipolar approach. 

We begin by analyzing the phonon distortion modes of BaCuF$_4$. As stated earlier, the CuF$_6$ rotational mode and the polar distortion mode are coupled to form the $\Gamma_2^-$ distortion mode, which is responsible for the proposed geometric ferroelectricity in BaCuF$_4$. For simplicity, we refer to this structure as the $+P$ structure.  
In the absence of these distortion modes, the structure becomes non-polar with space group symmetry $Cmcm$, as shown in Fig. \ref{fig3}(b). We refer to this hypothetical structure as the $P=0$ structure. Interestingly, by reversing the $\Gamma_2^-$ distortion and condensing it onto the $Cmcm$ structure, we obtain the polar $Cmc2_1$ structure with opposite polar displacements of the Ba atoms, as indicated in Fig. \ref{fig3}(c). We refer to this structure as the $-P$ structure, which is energetically degenerate with the $+P$ structure. 
As seen in Fig. \ref{fig3}(c), because of the coupling between the polar mode and the rotational distortion mode, when we switch from the $+P$ to the $-P$ structure using an external electric field \footnote{Considering the hypothetical $Cmcm$ structure as a reference structure, it was shown theoretically that the estimated energy barrier supports the possibility of achieving ferroelectric switching in practice in BaCuF$_4$. \cite{Garcia_BaCuF4_2018} 
} 
the rotation of the CuF$_6$ octahedra is also reversed along with the polar displacements.  
It is this octahedral rotation that plays the crucial role in tuning the NRSS, as we now discuss. 

\begin{table*}[htbp]
\centering
\small
\setlength{\tabcolsep}{4pt}
\renewcommand{\arraystretch}{1.1}
\caption{Computed ordering of relevant multipoles for the $+P$, $P=0$, and $-P$ structures (shown in Fig.~\ref{fig3}(a)–(c)) of BaCuF$_4$. The symbols $\pm$ denote the relative signs of a given multipole at the four Cu ions: Cu1 $(0,y,z)$, Cu2 $(0,-y,z+\tfrac{1}{2})$, Cu3 $(\tfrac{1}{2},y+\tfrac{1}{2},z)$, and Cu4 $(\tfrac{1}{2},-y+\tfrac{1}{2},z+\tfrac{1}{2})$ for the $+P$ and $-P$ structures. For the $P=0$ structure, the corresponding coordinates are $(0,y,\tfrac{1}{4})$, $(0,-y,\tfrac{3}{4})$, $(\tfrac{1}{2},y+\tfrac{1}{2},\tfrac{1}{4})$, $(\tfrac{1}{2},-y+\tfrac{1}{2},\tfrac{3}{4})$ respectively.}

\label{BaCuF4_mp}

\begin{tabular}{|c|c|c|c|c|c|c|c|}
\hline
\multirow{2}{*}{\textbf{Structure}} &
\multirow{2}{*}{\textbf{Multipoles}} &
\multirow{2}{*}{$\boldsymbol{w^{kpr}}$} &
\multicolumn{4}{|c|}{\textbf{Sign of multipoles at}} &
\multirow{2}{*}{\textbf{AF/F}} \\     
\cline{4-7}
& & &
\textbf{Cu1} &
\textbf{Cu2} &
\textbf{Cu3} &
\textbf{Cu4} & \\
\hline

\multirow{5}{*}{$+P$} &
Charge Quadrupole & $w^{202}_{-1}$ & $-$ & $+$ & $-$ & $+$ & AF \\
\cline{2-8}
& Magnetic Dipole  & $w^{101}_{0}$  & +  & --  & +  & --  & AF \\ 
\cline{2-8}
& \multirow{3}{*}{Magnetic Octupole} & $w^{211}_{-1}$ & $-$ & $-$ & $-$ & $-$ & F \\
&                    & $w^{212}_{1}$  & $+$ & $+$ & $+$ & $+$ & F \\
&                    & $w^{213}_{-1}$ & $-$ & $-$ & $-$ & $-$ & F \\
\hline

\multirow{5}{*}{$P=0$} &
Charge Quadrupole & $w^{202}_{-1}$ & 0 & 0 & 0 & 0 & -- \\
\cline{2-8}
& Magnetic Dipole  & $w^{101}_{0}$  & -- & + & --- & + & AF \\
\cline{2-8}
& \multirow{3}{*}{Magnetic Octupole} & $w^{211}_{-1}$ & 0 & 0 & 0 & 0 & -- \\
&                    & $w^{212}_{1}$  & 0 & 0 & 0 & 0 & -- \\
&                    & $w^{213}_{-1}$ & 0 & 0 & 0 & 0 & -- \\
\hline

\multirow{5}{*}{$-P$} &
Charge Quadrupole & $w^{202}_{-1}$ & $+$ & $-$ & $+$ & $-$ & AF \\
\cline{2-8}
& Magnetic Dipole  & $w^{101}_{0}$  & +  & --  & +  & --  & AF \\
\cline{2-8}
&\multirow{3}{*}{Magnetic Octupole} & $w^{211}_{-1}$ & $+$ & $+$ & $+$ & $+$ & F \\
&                    & $w^{212}_{1}$  & $-$ & $-$ & $-$ & $-$ & F \\
&                    & $w^{213}_{-1}$ & $+$ & $+$ & $+$ & $+$ & F \\
\hline
\end{tabular}
\end{table*}

The role of octahedral rotation becomes evident from our multipole calculations.  Our computed multipoles for the $+P$ structure with the Cu spins oriented along $\hat z$ show that the rotation of the CuF$_6$ octahedra gives rise to an antiferroic ordering of the charge quadrupole component $w^{202}_{-1}$ at the Cu sites. As listed in Table \ref{BaCuF4_mp}, interestingly, the antiferroic pattern of the charge quadrupole moment is identical to that of the AFM dipolar order at the Cu ions, resulting in a ferroic pattern of the magnetic octupole components $w^{211}_{-1}, w^{212}_{1}, \text{and } w^{213}_{-1}$. These magnetic octupoles describe an anisotropic magnetization density on the $yz$ plane with the spin polarization along the $\hat z$ direction, represented by the real-space representation $yzm_z$. The corresponding $k$-space representation of these magnetic octupole components is $k_yk_zm_z$ , which suggests the presence of NRSS along any direction in reciprocal space with $k_y\ne 0$ and $k_z\ne 0$, while the spin polarization of the bands is along the $\hat z$ direction. Consistent with this, our computed band structure for the $+P$ structure, shown in Fig. \ref{fig3}(d), exhibits NRSS along the $R (0, 0.5,0.5) \rightarrow \Gamma$ direction.

Based on our understanding of the multipolar origin of NRSS in the $+P$ structure, we expect that NRSS should be reversed in the $-P$ structure due to the reversal of the octahedral rotation, while it should be absent in the $P=0$ structure due to the absence of CuF$_6$ rotation. Motivated by this, we performed multipole analysis for both these structures, and our key findings are also listed in Table \ref{BaCuF4_mp}. For the $P=0$ structure, our multipolar analysis shows that the charge quadrupole component $w^{202}_{-1}$ at the Cu site vanishes, consistent with the absence of octahedral rotation. Consequently, even in the presence of the same AFM order, the absence of charge quadrupole components leads to the absence of ferroically ordered magnetic octupole components. This, in turn, results in the absence of NRSS in the $P=0$ structure, as also evident from our computed band structure in Fig. \ref{fig3}(e).  

More interestingly, the computed multipoles for the $-P$ structure, as listed in Table \ref{BaCuF4_mp}, show that the antiferroic charge quadrupole components $w^{202}_{-1}$ switch their signs while retaining the same antiferroic pattern as in the $+P$ structure. This can be understood from the reversal of the CuF$_6$ octahedral rotation in the $-P$ structure, as discussed earlier. Consequently, we find that the ferroically ordered magnetic octupole components $w^{211}_{-1}, w^{212}_{1}, \text{and } w^{213}_{-1}$ also switch their signs, which, in turn, reverses the NRSS in the band structure of the $-P$ structure compared to the $+P$ structure (see Fig. \ref{fig3} (f)). Thus, in ferroelectric materials hosting NRSS, if the polar distortion couples to the octahedral rotation, which is key to NRSS, switching the polar distortion using an external electric field will result in the reversal of NRSS, thereby offering the possibility of electric-field tuning of NRSS.

\subsection{Tuning of NRSS in Supperlattice Structures } 

\begin{figure}[t]
    \centering
\includegraphics[width=1\linewidth]{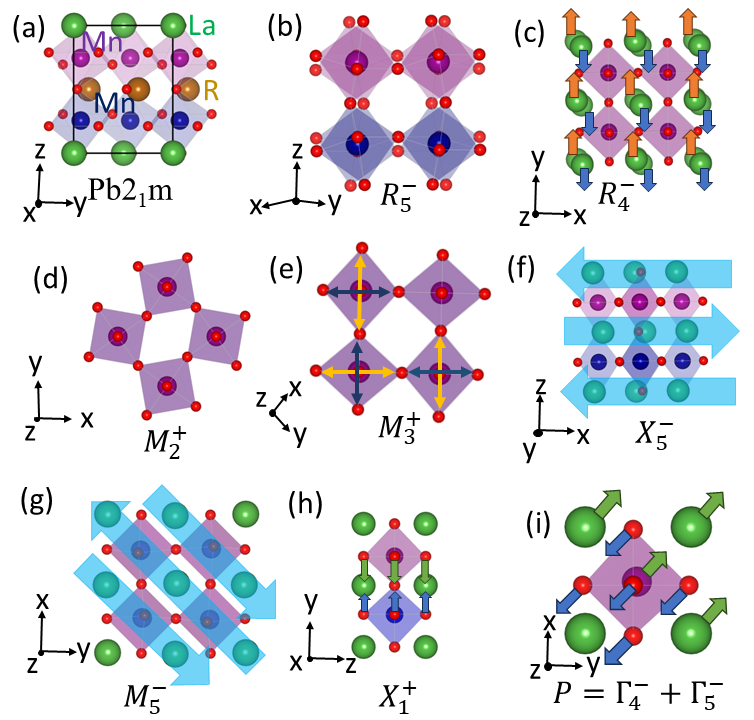}
    \caption{Atomic distortions of the LaMnO$_3$/$R$MnO$_3$ superlattice. (a) Crystal structure of LaMnO$_3$/$R$MnO$_3$ superlattice with space group symmetry $Pb2_1m$. Here, two different colors of Mn atoms represent opposite spin polarization.  (b)-(i) Different phonon distortion modes of $Pb2_1m$  superlattice  structure. Atomic distortions in (b)-(f) are also present in the $Pbnm$ structure of LaMnO$_3$, describing (b) out-of-phase octahderal rotation ($R_5^-$), (c) antipolar motion of La atoms ($R_4^-$) (d) in-phase octahderal rotation $M_2^+$, (e) Jahn-Teller distortion, ($M_3^+$) Here the arrows in blue color indicates the contraction and the arrows in brown color indicate the elongation and (f) antipolar motion of La atoms ($X_5^-$). Atomic distortions shown in (g)-(i) correspond to the additional distortion modes present in the $Pb2_1m$  superlattice  structures with the antipolar displacements of the $R$-site cations and O atoms in (g) $M_5^-$ and (h) $X_1^+$ distortions and (i) combined polar distortion $P$.}
    
    \label{fig4}
\end{figure}

Similar to the broken-inversion-symmetric case discussed above, even in an inversion-symmetric system, superlattice engineering can induce electric polarization, facilitating electric-field tuning of NRSS, as illustrated for the LaMnO$_3$/GdMnO$_3$ superlattice in our earlier work \cite{BandyopadhyayPRB2025}. LaMnO$_3$ crystallizes in the inversion-symmetric $Pbnm$ structure, and shows NRSS in its ground state with A-type AFM order. However, forming a LaMnO$_3$/GdMnO$_3$ superlattice breaks inversion symmetry, giving rise to {\it improper ferroelectricity} \cite{BENEDEK2012,Subhadeep_APL} with 
 a net in-plane electric polarization and thereby enabling electric-field tunability \cite{BandyopadhyayPRB2025}. Here, we demonstrate that, in addition to the electric-field-driven reversal of NRSS, it is also possible to tune the \emph{magnitude} of NRSS in LaMnO$_3$/$R$MnO$_3$ superlattices (see Fig. \ref{fig4} (a)) by substituting $R$ with different cations (Gd, Eu, Sm, and Nd).   

We begin by analyzing the different phonon modes of the $Pbnm$ structure of LaMnO$_3$ relative to the cubic $Pm\overline{3}m$ structure. The obtained five distortion modes are depicted in Figs.~\ref{fig4} (b)-(f). Among them, the dominant modes are $R_5^-$ and $M_2^+$, corresponding to anti-phase and in-phase rotations of the MnO$_6$ octahedra, respectively. The other three distortion modes are $X_5^-$, $M_3^+$, and $R_4^-$, which are associated with antipolar displacements of La atoms, Jahn-Teller distortions, and combined antipolar motions of La and O atoms, respectively. The LaMnO$_3$/$R$MnO$_3$ superlattice adopts the polar $Pb2_1m$ space group. Here, the mismatch between the ionic radii of La (1.045 \AA) and $R=$ Nd (0.983 \AA), Sm (0.958 \AA), Eu (0.947 \AA), and Gd (0.938 \AA) induces inequivalent octahedral rotations, resulting in uncompensated antipolar $X_5^-$ distortions and, in turn, a net electric polarization $\vec P$. Symmetry adapted  mode analysis reveals that, in addition to the $Pbnm$ modes, four new distortion modes, viz., $\Gamma_4^-$, $\Gamma_5^-$, $X_1^+$, and $M_5^-$ also appear in the $Pb2_1m$ superlattices (see Fig.~\ref{fig4} (g)-(i)). Among these, $\Gamma_4^-$ and $\Gamma_5^-$ are polar distortions that jointly contribute to the polarization and are, therefore, collectively denoted as $P$. On the other hand, $X_1^+$ and $M_5^-$ govern the antipolar motions of the $R$-site cations and O atoms  
as shown in Figs.~\ref{fig4} (g) and (h).  

Interestingly, the polar distortion mode $P$ couples trilinearly with other modes, giving rise to anharmonic terms in the free energy of the $Pb2_1m$ structure, 
\begin{equation} \label{FPb2_1m}
    F_{Pb2_1m} \propto  PM_2^+M_5^-
    + PM_3^+M_5^-.
\end{equation}
This trilinear coupling implies that modulation of the polar mode $P$ also changes the $M_2^+$ and $M_3^+$ distortion modes, which describe MnO$_6$ octahedral distortions. Since these octahedral distortions correlate with the magnitude of magnetic octupoles, controlling the polar distortion $P$ provides the means to tune both the octupole moments and the NRSS.    

\begin{figure}
    \centering
    \includegraphics[width=1\linewidth]{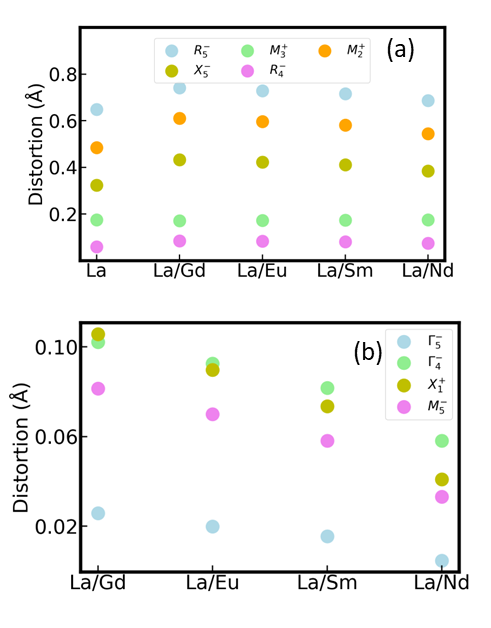}
    \caption{(a) Variation in the amplitude of different distortion modes, shown in Fig. \ref{fig4} (b)-(f) for LaMnO$_3$ and LaMnO$_3$/$R$MnO$_3$ superlattice structures. (b) The same variation for the atomic distortions shown in Fig. \ref{fig4} (g)-(i) for different superlattice structures.}
    \label{fig5}
\end{figure}

Motivated by this, we extracted the amplitudes of all distortions present in LaMnO$_3$/$R$MnO$_3$ superlattices ($R=$ Gd, Eu, Sm, and Nd) using symmetry adapted mode analysis (see Sec.~\ref{method} for computational details) and compared them with those of bulk LaMnO$_3$. The results are shown in Fig.~\ref{fig5}. As seen in the bottom panel, the amplitudes of the additional phonon modes, particularly the polar mode $P$, decrease as the size difference between La and the $R$ cation decreases. Thus, the amplitude of the polar distortion can be tuned systematically by varying the $R$-site cation.  

\begin{figure} 
  \centering
  \includegraphics[width=1\linewidth]{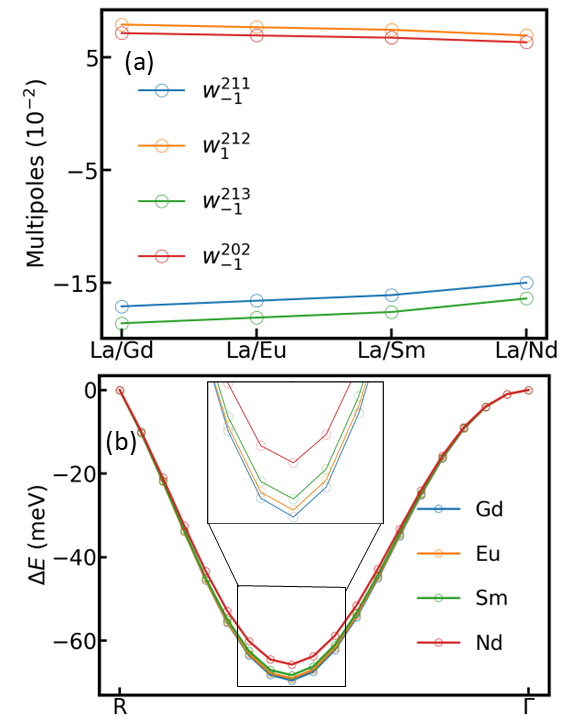}
  \caption{ Tuning of NRSS in LaMnO$_3$/$R$MnO$_3$ superlattice structures. (a) Variation of the antiferroic charge quadrupole component $w^{202}_{-1}$ (in units of electronic charge) and the ferroic magnetic octupole components $w^{211}_{-1}$, $w^{212}_{1}$, and $w^{213}_{-1}$ (in $\mu_B$) at the Mn site for different superlattices. (b) Variation of the energy difference $\Delta E$ between oppositely spin-polarized bands along R $\rightarrow$ $\Gamma$ for different superlattices. The inset shows a magnified view for clarity.
   }
  \label{fig6}
\end{figure}

To examine the effect of varying polar-mode distortions on magnetic multipoles and NRSS, we performed multipolar calculations for each superlattice structure and analyzed their corresponding band structures. For these calculations, we considered the A-type AFM order, the magnetic ground state of LaMnO$_3$, with spin polarization along the $\hat z$ direction.  Our results reveal the presence of an antiferroic ordering of the charge quadrupole component $w^{202}_{-1}$ with the same antiferroic pattern at the Mn ions as the A-type AFM order. As seen from Fig.~\ref{fig6}, we find that as the cation size mismatch increases, the magnitude of $w^{202}_{-1}$ also increases, while still maintaining the same antiferroic order. This trend can be attributed to the increase in $M_2^+$ and $M_3^+$ octahedral distortions accompanying the increase in the polar distortion amplitude, as shown in Fig.~\ref{fig5}. Consequently, the ferroically ordered magnetic octupole components \( w^{211}_{-1} \), \( w^{212}_{1} \), and \( w^{213}_{-1} \) at the Mn ions also increase in magnitude with larger cation size mismatch. This leads to an enhanced energy splitting $\Delta E$ between spin-up and spin-down bands, viz.,
$\Delta E = E_{\uparrow} - E_{\downarrow},$  
as evident from Fig.~\ref{fig6}(c). These results demonstrate that NRSS can be systematically tuned by substitution of the $R$ site with different cation in the LaMnO$_3$/$R$MnO$_3$ superlattice structure.

\section{Summary and outlook} 

To summarize, we demonstrate the tuning of NRSS in three example materials with diverse structural and magnetic symmetries. Despite these differences, we show that NRSS can be controlled in all cases by manipulating one or more atomic distortions that are crucial to the effect. As discussed in the manuscript, the identification of the most relevant atomic distortion can be guided by analyzing atomic-site multipoles, highlighting the role of higher-order multipoles in this class of materials.

Although the three example systems, considered here, exhibit $d$-wave NRSS, the same principle of controlling NRSS by tuning the relevant atomic distortion through external perturbations applies to other types of NRSS as well. For example, $g$-wave NRSS has been correlated with the magnetic triakontadipole \cite{Verbeek2024}, which depends on structural distortions that quantify the charge hexadecapole, in analogy with the magnetic octupole–charge quadrupole correlation discussed in this work. Thus, by manipulating distortions that govern the charge hexadecapole, we can tune the magnetic triakontadipole and, thereby, control the $g$-wave spin splitting.

We emphasize that the manipulation of NRSS discussed here is possible without affecting the N'eel vector. This highlights the role of the local structural environment in driving NRSS and the inherent structure–spin correlation in this family of materials. Indeed, this correlation acts as an effective substitute for spin–orbit interaction in lifting the spin degeneracy of electronic bands. We hope that our work will motivate further investigations into possible spin–phonon coupling in these systems.

Beyond the general framework for controlling NRSS, our work also provides material-specific insights. For example, our study of the effect of biaxial strain in MnF$_2$ complements previous investigations of shear strain in this material and its sister compound CoF$_2$, which revealed piezomagnetic behavior \cite{Dzialoshinskii1958,Borovikromanov1960,Baruchel1980,Baruchel1988}. A key distinction is that under biaxial strain, the symmetry of the unstrained structure is preserved, whereas shear strain breaks the symmetry, producing a net magnetic moment and, thus, contrasting with the compensated antiferromagnetic state in the former case. We hope that our theoretical predictions will stimulate future experiments exploring the role of biaxial strain in MnF$_2$.

Another important aspect concerns the detection of NRSS and its modulation in both magnitude (as shown for the LaMnO$_3$/RMnO$_3$ superlattice) and sign (e.g., in BaCuF$_4$). While detecting NRSS in insulating systems remains challenging, recent proposals suggest that it can be probed via shift-current measurements in inversion-symmetry-broken systems \cite{Gu2025}. We hope that our results will encourage experimental efforts in these directions.

\section*{Acknowledgements}
AR and SBh thank National Supercomputing Mission for providing computing resources
of ‘PARAM Porul’ at NIT Trichy, implemented by C-DAC and supported by the Ministry
of Electronics and Information Technology (MeitY) and Department of Science and
Technology, Government of India.
SBh gratefully acknowledges financial support from the IRCC Seed Grant (Project Code: RD/0523-IRCCSH0-018), the INSPIRE Research Grant (Project Code: RD/0124-DST0030-002), and the ANRF PMECRG Grant (Project Code: RD/0125-ANRF000-019).

\appendix

\section{Analysis of Phonon modes in the ground state structure of MnF$_2$} \label{appendix1}

\begin{table}[h]
\caption{Frequencies of the optical phonon modes in MnF$_2$ at the $\Gamma$ point. Modes marked with * are associated with the in-plane displacements of the F atoms. The last column indicates symmetry of the structures upon condensation of a particular distortion.}
\begin{tabular}{|c|c|c|}
\hline
& & \\
$~~~~$index$~~~~$ & $~~~~$frequency (cm$^{-1}$) $~~~~$ & $~~~~$ Symmetry $~~~~$\\
& & \\
\hline
\hline
 & & \\
$~$4$^*$ ($B_{1g}$) & 63 & $Pnnm$\\
\hline
& & \\
5 & 119 & $P\overline{4}2_1m$\\
\hline
& & \\
6 & 157 & $Pm$\\
\hline
& & \\
7 & 157 & $Pm$\\
\hline
& & \\
$~$8$^*$ ($A_{2g}$) & 220  & $P4_2/m$\\
\hline
& & \\
9 & 234   & $C2/m$\\
\hline
& & \\
10 & 234  & $C2/m$\\
\hline
& & \\
11 & 247 & $Pm$\\
\hline
& & \\
12 & 247 & $Pm$\\
\hline
& & \\
13 & 284  & $P4_2nm$ \\
\hline
& & \\
14 & 318  & $P\overline{4}2_1m$ \\
\hline
& & \\
$~$15$^*$ ($A_{1g}$) & 325  & $P4_2/mnm$ \\
\hline
& & \\
16 & 348  & $Pm$ \\
\hline
& & \\
17 & 348  & $Pm$\\
\hline
& & \\
$~$18$^*$ ($B_{2g}$) & 453 & $Cmmm$\\
\hline
\end{tabular}
\label{tab2}
\end{table} 

\begin{figure} [h]
    \centering
    \includegraphics[width=1\linewidth]{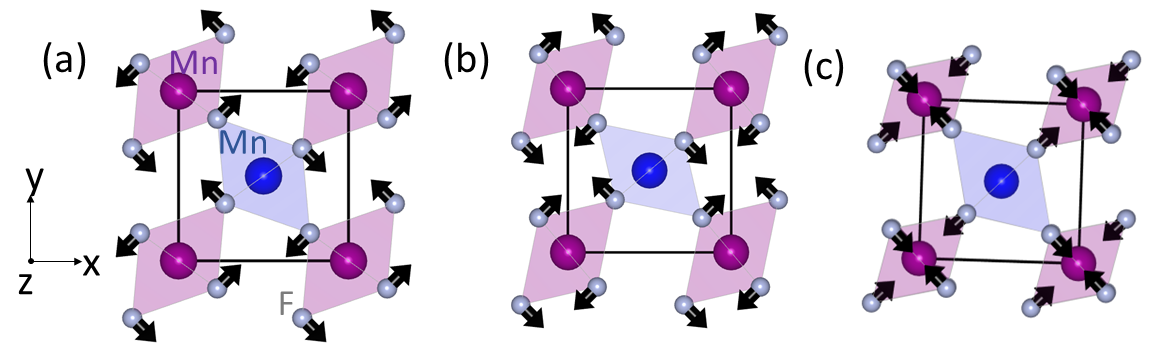}
    \caption{Atomic distortions of MnF$_2$, showing the in-plane displacements of the F atoms. (a) B$_{1g}$, (b) A$_{2g}$, and (c) B$_{2g}$ phonon modes. The black arrows indicate the displacements of the F atoms.}
    \label{fig7}
\end{figure}

To analyze the different atomic distortions, we computed the phonon band structure of MnF$_2$ using the finite difference method as implemented in PHONOPY, considering the ground state AFM order in MnF$_2$ as depicted in Fig. \ref{fig1}b~\cite{Phonopy}. Since the unit cell of MnF$_2$ has six atoms, there are in total 18 phonon branches present in the phonon band structure. Our calculation shows that no unstable phonon mode is present, confirming the dynamic stability of the ground state $P4_2/mnm$ structure of MnF$_2$. 

The computed frequencies of the 15 optical phonon modes at the $\Gamma$ point are listed in Table \ref{tab2}. Out of these 15 stable optical phonon modes, those that describe displacements of the F atoms on the $a-b$ plane are marked in Table \ref{tab2} and are also depicted schematically at the inset of Fig. \ref{fig2} and in Fig \ref{fig7}. These in-plane atomic distortions can possibly impact the charge quadrupole component $w^{202}_{-2}$, describing anisotropic charge density on the $xy$ plane. It is this $w^{202}_{-2}$ component that couples to the AFM dipolar order to give rise to the ferroically ordered magnetic octupoles in MnF$_2$. Interestingly, as evident from Table \ref{tab2}, only one of these modes (mode 15 in Table \ref{tab2}) preserves the $P4_2/mnm$ symmetry of the MnF$_2$ ground state structure upon condensation, while others do not. This mode 15 at frequency 325 cm$^{-1}$, known as $A_{1g}$ phonon mode, involves antiferroic distortion of the MnF$_6$ networks as shown schematically in the inset of Fig. \ref{fig2}. Since this atomic distortion is consistent with the antiferroic charge quadrupole moment and preserves the ground state symmetry of the MnF$_2$ structure, we focus on manipulating the $A_{1g}$ phonon mode, as discussed in section \ref{MnF2}.

\bibliography{main}
\end{document}